# Naming the Pain in Machine Learning-Enabled Systems Engineering

Marcos Kalinowski[a], Daniel Mendez[b], Görkem Giray[c], Antonio Pedro Santos Alves[a], Kelly Azevedo[a], Tatiana Escovedo[a], Hugo Villamizar[a], Helio Lopes[a], Teresa Baldassarre[d], Stefan Wagner[e], Stefan Biffl[f], Jürgen Musil[f], Michael Felderer[g,h], Niklas Lavesson[b], Tony Gorschek[b]

[a]*Pontifical Catholic University of Rio de Janeiro (PUC-Rio), Brazil*
[b]*Blekinge Institute of Technology (BTH), Sweden*
[c]*Independent Researcher, Turkey*
[d]*University of Bari, Italy*
[e]*Technical University of Munich (TUM), Germany*
[f]*Vienna University of Technology (TUW), Austria*
[g]*German Aerospace Center (DLR), Germany*
[h]*University of Cologne, Germany*

---

## Abstract

**Context:** Machine learning (ML)-enabled systems are being increasingly adopted by companies aiming to enhance their products and operational processes.

**Objective:** This paper aims to deliver a comprehensive overview of the current status quo of engineering ML-enabled systems and lay the foundation to steer practically relevant and problem-driven academic research.

**Method:** We conducted an international survey to collect insights from practitioners on the current practices and problems in engineering ML-enabled systems. We received 188 complete responses from 25 countries. We conducted quantitative statistical analyses on contemporary practices using bootstrapping with confidence intervals and qualitative analyses on the reported problems using open and axial coding procedures.

**Results:** Our survey results reinforce and extend existing empirical evidence on engineering ML-enabled systems, providing additional insights into typical ML-enabled systems project contexts, the perceived relevance and complexity of ML life cycle phases, and current practices related to problem understanding, model deployment, and model monitoring. Furthermore, the qualitative analysis provides a detailed map of the problems practitioners face within each ML life cycle phase and the problems causing overall project failure.

**Conclusions:** The results contribute to a better understanding of the status quo and problems in practical environments. We advocate for the further adaptation and dissemination of software engineering practices to enhance the engineering of ML-enabled systems.

---

## 1. Introduction

Companies from different sectors are increasingly incorporating machine learning (ML) components into their software systems. We refer to these software systems, where an ML component is part of a larger system, as ML-enabled systems. The shift from engineering conventional software systems to ML-enabled systems comes with challenges related to the idiosyncrasies of such systems, such as the high dependency on data, addressing additional qualities properties (*e.g.,* fairness and explainability), dealing with iterative experimentation, and facing unrealistic assumptions [31, 25]. In fact, the non-deterministic nature of ML-enabled systems poses software engineering (SE) challenges [7] and there are many specific concerns to be considered when engineering ML-enabled systems [32].

Mature tools and techniques for engineering ML-enabled systems are still missing [7], and SE can play an important role in resolving issues associated with the development of these systems. For instance, the literature suggests that requirements engineering (RE) can help to address problems related to dealing with customer expectations and to better aligning requirements with data [33, 31, 1]. Furthermore, designing and developing ML-enabled systems is complex and can be eased by having a good software architecture and making effective design decisions [26]. Indeed, many challenges of ML-enabled systems can be addressed from a software architecture perspective [19, 25, 26], and software architecture is also a cornerstone for effective production deployment of such systems [37].

Understanding the status quo is fundamental for properly adapting SE practices for this context. Considering this, we conducted an international survey to better understand the current industrial practices and problems practitioners face when engineering ML-enabled systems. In total, 188 practitioners from 25 countries completely answered the survey. We conducted quantitative and qualitative analyses based on practitioners' responses, allowing us to gain insights into practices and problems.

In our previous work, we reported the findings related to RE [1]. In this extension, we report on the complete survey findings, strengthening the empirical evidence on (i) typical ML-enabled systems project contexts, (ii) perceived relevance and complexity of each ML life cycle phase, (iii) contemporary practices of RE for ML-enabled systems, (iv) contemporary practices for deploying and monitoring ML models, and (v) the main problems reported by practitioners on engineering ML-enabled systems. Sharing such findings on the state of practice and problems with the community shall help steer SE research toward contributing to better ML-enabled systems engineering. We briefly summarize the main findings hereafter.

Our findings indicate that ML-enabled system projects are mostly managed using agile practices, with the Scrum and Kanban frameworks being particularly popular in this context. Still, more than 30% do not use any kind of management framework. Most of these projects use ML for classification or regression problems. Many different ML algorithms have been used in these projects, with neural networks being the most popular, followed closely by decision trees and ensembles combining multiple models.

Concerning the relevance and complexity of the ML life cycle phases, all phases are perceived as highly relevant, with problem understanding being the most relevant, followed by data collection, pre-processing, and model evaluation. Regarding the perceived complexity, problem understanding also takes the lead, followed by data collection, pre-processing, and model deployment. Problem understanding, data collection, and pre-processing are the phases that typically require the most effort.

With respect to the contemporary practices of RE for ML-enabled systems, we found significant differences. For instance, RE-related activities are mostly conducted by project leaders and data scientists, and the prevalent requirements documentation format concerns interactive notebooks. The main focus of non-functional requirements includes data quality, model reliability, and model explainability, and the main challenges include managing customer expectations and aligning requirements with data.

On the contemporary practices for deploying and monitoring ML models, models are mainly deployed as separate services. Embedding the model within the consuming application and platform-as-a-service solutions are less frequently explored. We also observed limited adoption of MLOps principles, with many projects not having an automated pipeline to retrain and redeploy the models. Many models in production are not monitored at all, and when monitored, the main monitored aspects are inputs, outputs, and decisions.

Finally, we organized the information on the main problems emerging from the practitioners' responses. Therefore, we applied open and axial coding procedures and used probabilistic cause-effect diagrams [14] to represent the results graphically. These are basically fishbone diagrams showing the frequency with which the different problems were mentioned by the participants. We built one of these for each ML life cycle phase and one for problems leading to project failure. For instance, for the problem-understanding phase, the qualitative analyses revealed that ML-enabled system projects tend to have unclear goals and requirements, that practitioners commonly lack business domain understanding, and that they face low customer engagement and communication issues. The problems revealed for each ML life cycle phase will be presented in detail throughout the paper. Furthermore, issues with problem understanding, expectation management, and data quality are perceived as commonly leading to project failure.

The remainder of this paper is organized as follows. Section II provides the background and related work. Section III describes the research method. Section IV presents the survey results. Sections V and VI discuss the results and threats to validity. Finally, Section VII presents our concluding remarks.

## 2. Background and Related Work

### *2.1. ML Life Cycle Phases*

ML encompasses a series of interconnected phases that collectively form what is commonly referred to as the ML life cycle. We used the abstraction of seven generic ML life cycle phases adopted by a Brazilian textbook on SE for data science [12], which was adapted from the CRISP-DM industry-independent process model phases [28]: *Problem Understanding and Requirements, Data Collection, Data Pre-Processing, Model Creation and Training, Model Evaluation, Model Deployment,* and *Model Monitoring*. These phases are shown in Figure 1.

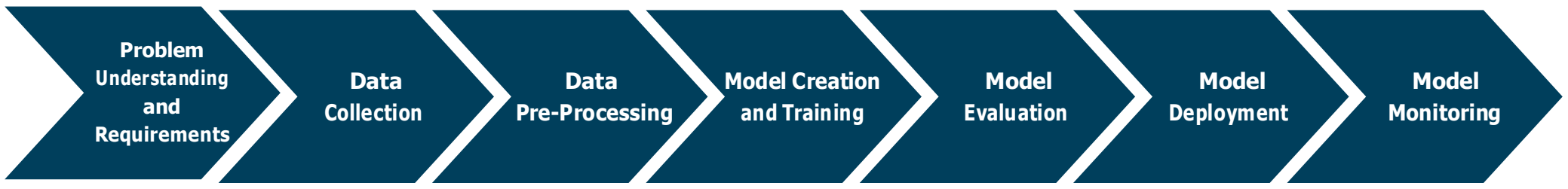

Figure 1: Seven ML life cycle phases considered in our study.

These phases directly relate to the nine ML life cycle phases presented by Amershi *et al.* [2] that have been commonly used as a reference for SE studies on ML-enabled systems. The only difference is that their phases of *Data Cleaning, Data Labeling*, and *Feature Engineering* in our chosen representation were cohesively grouped into a more generic Data Pre-Processing phase. This allowed our survey to be more flexible, coping with potentially different ways of pre-processing data. For instance, data labeling is only required for supervised learning, and specific contexts may require other pre-processing techniques, such as data transformations, augmentation, etc. A summarized description of the life cycle phases, based on [12], follows.

*Problem Understanding and Requirements.* This phase involves defining the problem the ML-enabled system aims to solve. It requires understanding the domain, identifying stakeholders' needs, and articulating precise project objectives. This stage sets the scope and direction of the entire project, requiring detailed documentation of requirements and expected outcomes.

*Data Collection.* This phase involves gathering relevant data from various sources. The quality and quantity of data collected directly impact the performance of the ML model. This phase must address data relevance, privacy issues, and ethical considerations, ensuring that the data is representative of the problem domain.

*Data Pre-processing.* Once data is collected, it often requires cleaning and transformation to make it suitable for modeling. Data pre-processing may include removing inconsistent data and outliers, handling missing values, data labeling, encoding categorical variables, normalizing data, and feature selection. This phase can be crucial for enhancing model accuracy and efficiency.

*Model Creation and Training.* In this phase, algorithms are selected and applied to the prepared data to create the ML model. Model creation involves choosing appropriate ML algorithms and configuring them with initial parameters. Training the model requires a subset of data where the model learns to predict outcomes effectively.

*Model Evaluation.* After training, the model is evaluated using a separate dataset not seen by the model during training. This phase assesses the model's performance against specific metrics, depending on the problem type. Model evaluation aims to ensure that the model generalizes well to new data.

*Model Deployment.* Deployment involves integrating the ML model into the existing production environment, where it can make real-time predictions. This phase requires careful planning to handle infrastructure needs, scaling, and integration with existing systems.

*Model Monitoring.* The final phase of the life cycle concerns monitoring the deployed model to ensure it performs as expected over time, identifying potential performance decays. Monitoring is crucial for maintaining the reliability and relevance of the model over time.

### *2.2. Challenges in Engineering ML-Enabled Systems*

Unlike traditional systems, ML-enabled systems learn from data instead of being programmed with predefined rules. The characteristics of such systems, including the high dependency on data and their non-deterministic nature, pose particular challenges to SE [7]. This section reports on studies exploring practices and challenges related to engineering ML-enabled systems, reflecting on the complexities introduced by ML components. To provide an overview and position our research, we organize some relevant related studies in chronological order into three research strategies: gathering experiences from industry engagements (*e.g.*, case studies, practitioner interviews), conducting literature-based studies (e.g., secondary studies and meta-summaries), and conducting large-scale questionnaire surveys.

*Experiences from Industry Engagements.* The study by Arpteg *et al.* [3] delved into SE challenges specific to deep learning, identifying twelve principal challenges across development, production, and organizational spectrums through interpretive research with various companies. Lwakatare *et al.* [22] proposed a taxonomy of challenges through case studies across different domains, enhancing understanding of the integration of ML in software-intensive systems. Amershi *et al.* [2] conducted a case study on SE for ML observing software teams at Microsoft. They identified typical ML life cycle phases and reported challenges related to data issues, ML component reuse, and modularization. Lewis *et al.* [19] focused on software architecture, providing four categories of architectural challenges collected via workshops and practitioner interviews: ML-enabled systems software architecture practices, architecture patterns and tactics for ML quality attributes, monitorability, and co-architecting and co-versioning.

*Literature-Based Studies.* Lwatakare *et al.* [23] conducted a systematic literature review focusing on the development and maintenance of ML-based software systems. Their study identified 23 challenges and 8 solutions related to adaptability, scalability, safety, and privacy, emphasizing the difficulties in adapting and scaling ML workflows from data acquisition to deployment. Giray [7] systematically synthesized the current state of SE research for engineering ML systems through a systematic literature review. He identified a focus on software testing and that none of the SE for ML proposals had a mature set of tools and techniques. Martínez-Fernández *et al.* [24] conducted a systematic mapping study to understand SE for artificial intelligence-based systems. They also identified the prevalence of studies on testing and neglected areas, such as maintenance. They found that data-related issues are the most recurrent challenges. Both studies emphasize the need for continued exploration and development of SE practices for ML-enabled systems. Nahar *et al.* [25] conducted a systematic literature survey to consolidate knowledge about challenges in building software products with ML components based on existing studies that interviewed or surveyed industry practitioners across multiple projects. They categorize these challenges into the following areas: Requirements Engineering; Architecture, Design, and Implementation; Model Deployment; Data Engineering; Quality Assurance; Process; and Organization. Nazir *et al.* [26] combined a systematic literature review with expert interviews to further explore and categorize challenges and best practices focused on the architecture of ML-enabled systems.

*Large-Scale Surveys.* Ishikawa and Yoshioka [8] reported in 2019 results of a large-scale questionnaire-based survey created by a Japanese academic society (JSSST-MLSE) that gathered answers from 278 individuals who have worked on ML or ML applications in practice in Japan. The survey covered experiences with SE activities and ML techniques, quality attributes, difficulty levels of activities related to engineering ML-enabled systems, and characteristics of ML that lead to such difficulties. They found that existing techniques were, to a great perceived extent, not applicable anymore or that methods and tools were immature for typical engineering activities of such systems (in particular, decision-making with customers and software testing). Out of a list of pre-defined difficulties stemming from the specific nature of ML, imperfection, lack of oracle, and uncertainty of the implemented behavior were most frequently selected. According to the authors, this survey was intentionally designed mainly with predefined categories to grasp the over-all trend of difficulties and did not include a deep qualitative investigation into each item. Nevertheless, they had an open-text question on the reasons behind the previous answers, which received 127 answers and mainly helped explain the difficulty levels the participants gave to the different activities.

Wan *et al.* [36] performed a mixture of qualitative and quantitative studies with 14 interviewees and 342 survey respondents. Their study, also reported in 2019, focused on eliciting significant differences between ML and non-ML systems development. The results uncovered significant differences in SE activities (*e.g.*, requirements, design, testing, and process). In their study design, the purpose of the survey was to gather opinions to quantitatively validate statements on the differences uncovered in the interviews through a 5-point Likert scale. They found that the differences mainly originate from inherent ML features, such as uncertainty and the data for use. The detailed list of validated differences provides important insights for the community.

We complement these valuable studies with additional empirical evidence on current practices and problems related to engineering ML-enabled systems obtained from an international survey with practitioners. We differentiate from previous survey efforts by timing and purpose. Rather than focusing on difficulty levels and differences, we characterize the status quo and problems of engineering ML-enabled systems. Therefore, we designed the survey with an international team of researchers, grounding the questions on practices on the recent literature and letting the problems emerge from qualitative investigations based on open-text responses from the practitioners. Following best practices for survey research [35], we applied inferential statistics using bootstrapping with confidence intervals to characterize the status quo and conducted open and axial coding procedures from grounded theory to identify the main problems. As a result, we validate and extend existing findings, unveiling novel practical insights that can help guide current practices and future research in the field.

## 3. Research Method

### *3.1. Goal and Research Questions*

This paper aims to characterize the contemporary practices and problems experienced by practitioners when engineering ML-enabled systems. In particular, we want to characterize the project context (*e.g.*, management approaches, business sectors, kind of ML problems, ML algorithms, pro-gramming languages), the relevance and complexity of each ML life cycle phase, and, mainly, understand the contemporary practices and problems.

For contemporary practices, emphasizing the SE perspective, we want to dig deeper into the practices related to the problem understanding, model deployment, and model monitoring phases, as requirements and infrastructure-related aspects directly affect the engineering of the ML component and of the solution architecture. Hence, to keep the survey focused, after internal discussions, we decided to focus the contemporary practices questions on these phases instead of phases more closely related to specific aspects of data engineering or data science (*e.g.*, data collection, data pre-processing, model creation, and model evaluation).

For the problems, on the other hand, we want to identify the main pain points of engineering ML-enabled systems, considering all of its typical activities. Therefore, we decided to qualitatively survey the participants on problems related to each ML life cycle phase and on problems most likely leading to project failure. Given this scenario, we defined the following research questions:

- **RQ1. What is the typical project context of ML-enabled systems?** This question aims to characterize these projects in terms of how they are managed, the kind of ML problems they address, ML algorithms, and programming languages.
- **RQ2. What is the perceived relevance and complexity of the ML life cycle phases?** This question aims to characterize the perceived relevance and complexity of the different ML life cycle phases using Likert scales. We also discuss the perceived effort distribution among these phases, as complexity and effort are commonly interpreted in an intertwined way.
- **RQ3. What are the contemporary practices of RE for ML-enabled systems?** This question aims to reveal how practitioners are currently approaching RE for ML, identifying prevalent methods and the extent to which the industry aligns with established practices. We refined *RQ3* into more detailed questions as follows:
  - RQ3.1 Who is addressing the requirements of ML-enabled system projects?

- RQ3.2 How are requirements typically elicited in ML-enabled system projects?
  - RQ3.3 How are requirements typically documented in ML-enabled system projects?
  - RQ3.4 Which non-functional requirements (NFRs) typically play a major role in ML-enabled system projects?
  - RQ3.5 Which activities are considered to be most difficult when defining requirements for ML-enabled system projects?

- **RQ4. What are contemporary practices for deploying and monitoring ML models?** Similarly to the previous question, this one aims to reveal the practices and trends of the deployment and monitoring phases. We refined *RQ4* into more detailed questions as follows:
  - RQ4.1. Which approaches are used to deploy ML models?
  - RQ4.2. What are the MLOps practices and principles used?
  - RQ4.3. What percentage of projects deployed into production have their ML models being monitored?
  - RQ4.4. Which aspects of the ML models are monitored?

- **RQ5. What are the main problems reported by practitioners in ML-enabled system projects?** This question aimed at identi- fying the main problems for each ML life cycle phase and the ones most likely leading to overall project failure. We applied open and axial coding procedures to allow the problems to emerge from open-text responses provided by the practitioners. We refined RQ5 into the following questions:
  - RQ5.1. What are the main problems reported for each ML life cycle phase?
  - RQ5.2. What problems were reported as leading to overall ML-enabled system project failure?

### *3.2. Survey Design*

We designed our survey based on best practices of survey research [35], carefully conducting the following steps:

- **Step 1. Initial Survey Design**. We designed the survey instrument based on our research questions. It included a consent form, demographic questions on the participants and their ML-enabled system project context, and substantive questions focused on the ML life cycle and its problems, requirements, and model deployment and monitoring. While the substantive questions on the problems were open-ended, we relied on the literature to provide the theoretical foundations for questions and answer options on requirements [5, 34, 31], and model deployment and monitoring [10, 9, 27]. Therefrom, the initial survey was drafted by SE and ML researchers of PUC-Rio (Brazil) with experience in R&D projects involving deliveries of ML-enabled systems.

- **Step 2. Survey Design Review**. The survey was reviewed and adjusted based on online discussions and annotated feedback from SE and ML researchers of BTH (Sweden). Thereafter, the survey was also reviewed by the other co-authors.

- **Step 3. Pilot Face Validity Evaluation**. This evaluation involved a lightweight review by randomly chosen respondents. It was conducted with 18 Ph.D. students taking a Survey Research Methods course at UCLM (Spain) (taught by the first author). They were asked to provide feedback on the clearness of the questions and to record their response time. This phase resulted in minor adjustments related to usability aspects and unclear wording. The answers were discarded before launching the survey.

- **Step 4. Pilot Content Validity Evaluation**. This evaluation involved subject experts from the target population. Therefore, we selected five data scientists experienced in developing ML-enabled systems, asked them to answer the survey, and gathered their feedback. The participants had no difficulties answering the survey, which took an average of 20 minutes. After this step, the survey was considered ready to be launched.

The final survey started with a consent form describing the purpose of the study and stating that it was conducted anonymously. The remainder was divided into 15 demographic questions (D1 to D15) and three specific parts with 17 substantive questions (Q1 to Q17): seven on the ML life cycle and problems, five on requirements, and five on deployment and monitoring.

The survey was implemented using the Unipark Enterprise Feedback Suite. The complete final survey instrument can be found in our online open science repository [15].

### *3.3. Data Collection*

Our target population concerns professionals involved in engineering ML-enabled systems, including different activities, such as management, design, and development. Therefore, it includes practitioners in positions such as project leaders, requirements engineers, data scientists, and developers. We used convenience sampling, sending the survey link to professionals active in our partner companies, and also distributed it openly on social media. We excluded participants who informed that they had no experience with ML-enabled system projects. Data collection was open from January 2022 to April 2022. We received responses from 276 professionals, of which 188 completed all four survey sections. The average time to complete the survey was 21 minutes. While the survey was divided into four parts, we conservatively considered only data from the 188 fully completed survey responses in all our analyses.

### *3.4. Data Analysis Procedures*

For data analysis purposes, given that all questions were optional, the number of responses out of the 188 participants who completed the survey varies across the questions. Therefore, we explicitly indicate the number of responses $N$ when analyzing each question.

We used inferential statistics to analyze the closed questions on the project context and contemporary practices. Our population has an unknown theoretical distribution (*i.e.*, the distribution of ML-enabled system professionals is unknown). In such cases, resampling methods, like bootstrapping, have been reported to be more reliable and accurate than inference statistics drawn directly from the samples [21, 35]. Hence, we use bootstrapping to calculate confidence intervals for our results, similar as done in [34]. In short, bootstrapping involves repeatedly resampling with replacements from the original sample and calculating the statistics based on these resamples. For each question, we take the sample of $N$ responses for that question and bootstrap $S$ resamples (with replacements) of the same size $N$ . Following advice from the field of statistics, we set $N$ as the total valid answers of each question [4], and we set $S$ as 1000, which is a value that is reported to allow meaningful statistics [18]. The employed bootstrapping strategy is illustrated in Figure

2. The analysis scripts were implemented by the fourth author and validated by the first author.

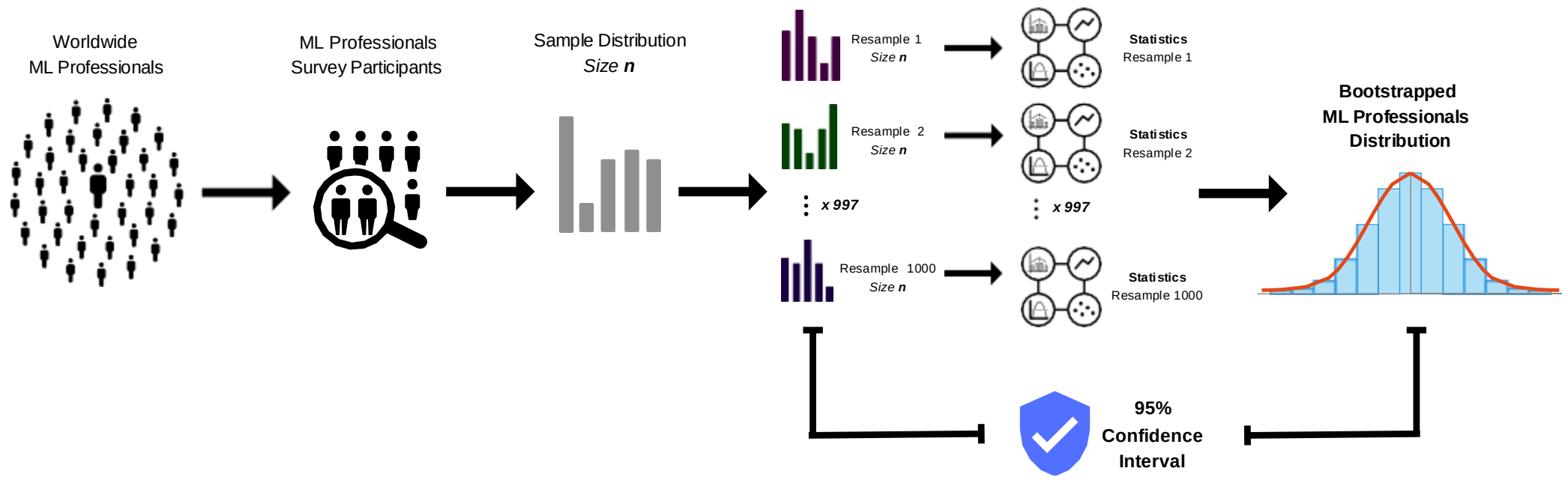

Figure 2: Bootstrapping with confidence intervals strategy.

For the open-text questions on the main problems faced by practitioners involved in engineering ML-enabled, we conducted a qualitative analysis using open and axial coding procedures from grounded theory [29]. This approach allows the problems to emerge from the open-text responses reflecting the experience of the practitioners. The qualitative coding procedures were conducted by the fifth author and initially reviewed by the first author (both from Brazil). Thereafter, the coding was reviewed independently at separate sites by the second and third authors (from Sweden and Turkey, respectively).

The questionnaire, the collected raw data, and the quantitative and qualitative data analysis artifacts, including Python scripts for the bootstrapping statistics and graphs and the peer-reviewed qualitative coding spreadsheets, are available in our online open science repository [15].

## 4. Survey Results

### *4.1. Study Population.*

Figure 3 summarizes demographic information on the survey participants' countries, roles, and experience with ML-enabled system projects in years. It is possible to observe that the participants came from different parts of the world, representing various roles and experiences. While the figure shows only the twelve countries with the most responses, we had respondents from 25 countries. As expected, the convenience sampling strategy of directly approaching the industry partners of the authors influenced the countries, with

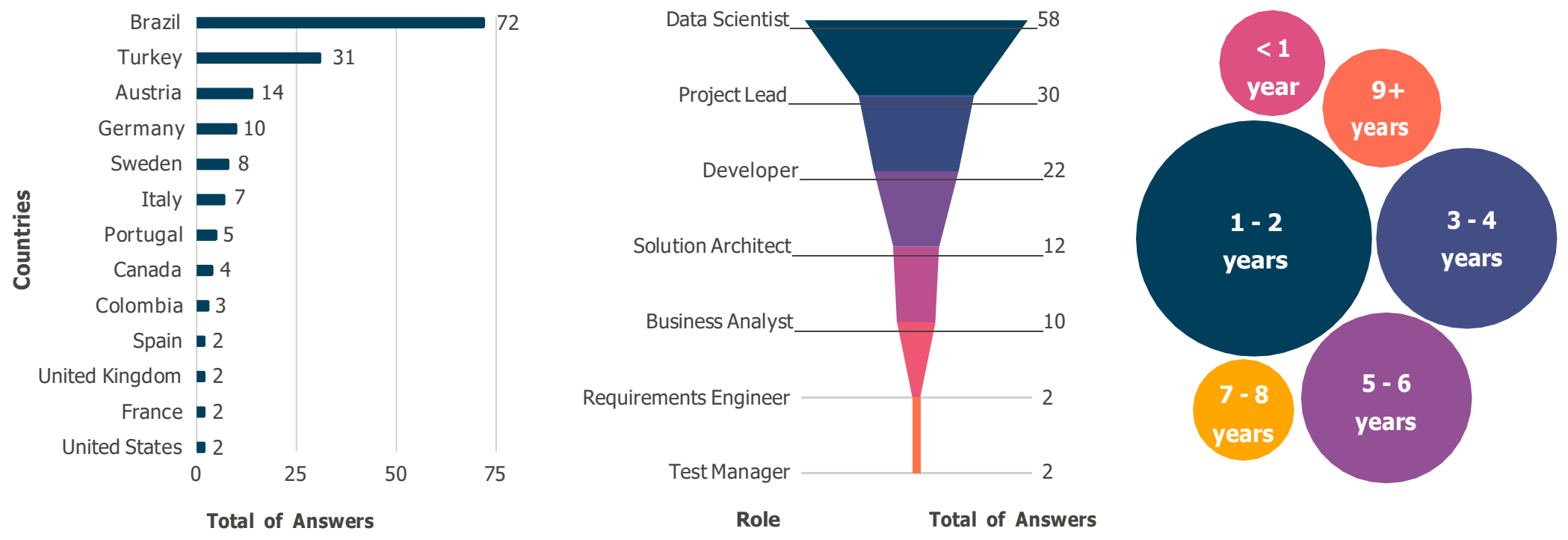

(a) Participants' Demographics (N = 175) (b) Participants' Main Role (N = 177) (c) Participants' ML Experience (N = 176)

Figure 3: Demographics: countries, roles, and years of experience.

most responses being from the authors' countries (Brazil, Turkey, Austria, Germany, Sweden, and Italy).

Regarding employment, 45% of the participants are employed in large companies (2000+ employees), while 55% work in smaller ones of different sizes. It is possible to observe that they are mainly data scientists, followed by project leaders, developers, and solution architects. Regarding their experience with ML-enabled systems, most of the participants reported having 1 to 2 years of experience. Following closely, another substantial group of participants indicated a higher experience bracket of 3 to 6 years. This distribution highlights a balanced representation of novice and experienced practitioners. Regarding the participants' educational background, 81% mentioned having a bachelor's degree in computer science, electrical engineering, information systems, mathematics, or statistics. Moreover, 54% held master's degrees, and 23% completed Ph.D. programs.

Furthermore, participants reported developing ML-enabled systems for various different business sectors, as depicted in the word cloud in Figure 4. The most frequent business sectors were Banking, Healthcare, Logistics, Sales, Telecom, Oil, Education, and Defense. This indicates the broad applicability of ML-enabled systems across industries and highlights the diversity of our sample.

### *4.2. RQ1. ML-Enabled Systems Project Contexts*

We characterize the projects in terms of their management, the kind of ML problems they address, ML algorithms, and programming languages. As we used bootstrapping, we hereafter report the mean bootstrapped proportion **P** of the participants that selected each answer and its 95% confidence interval in square brackets. To improve readability, the bootstrapping results are presented selectively to reinforce the interpretation of the results. The bootstrapping results for each answer option of each question can be found in the scripts contained in our online open science repository [15].

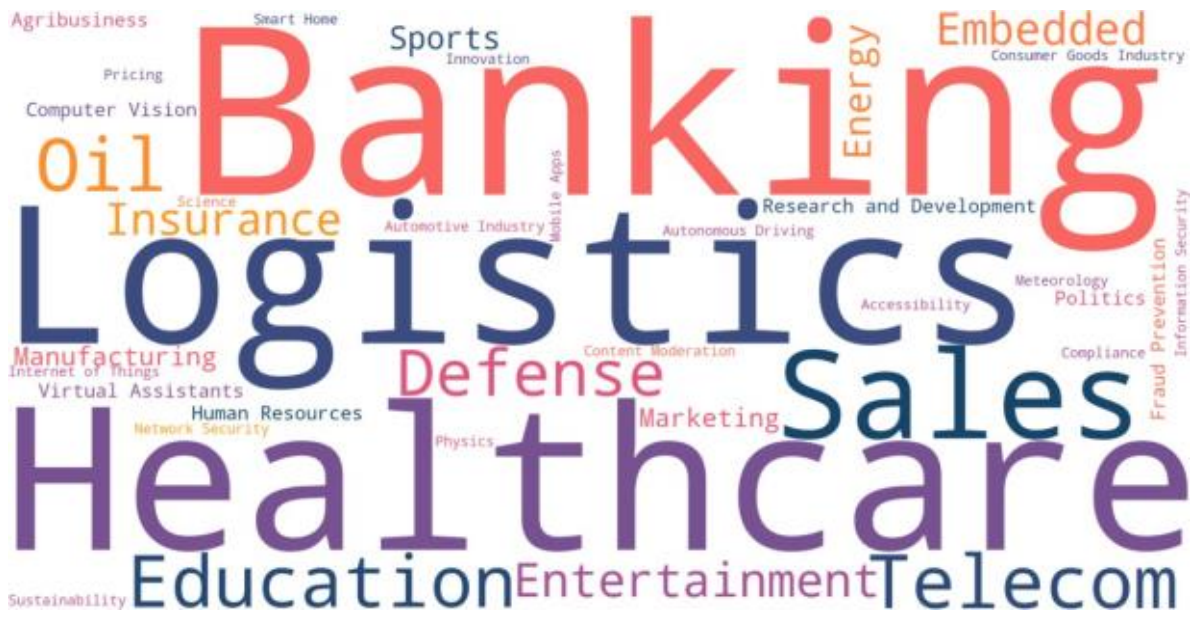

Figure 4: Participants' business sector (N = 175).

The proportion of participants involved in projects using each project management framework is presented in Figure 5 (a) together with the 95% confidence interval. It is possible to observe that Scrum is by far the most used project management framework for engineering ML-enabled systems (**P = 53.588 [53.341, 53.835]**). Kanban is also relatively popular in this context (**P = 33.352 [33.122, 33.582]**). However, it is also noteworthy that more than 30% (**P = 33.033 [32.772, 33.294]**) do not use any management framework. Some isolated options were mentioned in the "Others" field (*e.g.*, task lists, V Model, and tailored waterfall), altogether summing up less than 4% and not significantly influencing the overall distribution (**P = 3.648 [3.559, 3.737]**).

On the perceived degree of agility, participants reported the agility of their development as mostly agile, as presented in Figure 5 (b). It is possible to observe the highest bootstrapping proportion falling under mostly agile option (**P = 34.685 [34.425, 34.945]**), followed closely by balanced approaches (**P = 32.082 [31.884, 32.28]**).

The kinds of ML problems being handled by the ML-enabled systems are presented in Figure 5 (c). It is possible to observe that classification (**P = 74.156 [73.953, 74.358]**) and regression (**P = 73.247 [73.045, 73.449]**) are the most common types of problems being addressed. Cluster-ing is also significant (**P = 44.483 [44.245, 44.721]**). Some other isolated topics related to computer vision (*e.g.*, pose detection and estimation, image segmentation, image translation) and natural language processing were also mentioned in the "Others" field.

The main used ML algorithms are reported in Figure 5 (d). It is possible to observe a variety of algorithms being used in ML-enabled system projects, with neural networks (**P = 59.678 [59.453, 59.904]**), decision trees (**P = 56.014 [55.76, 56.268]**), and ensembles combining several models (**P = 45.437 [45.199, 45.674]**) being the most used ones. When analyzing this data, it is noteworthy to remember that there is a relation between the algorithms and the types of problems. For instance, KMeans is an algorithm specifically designed for clustering tasks. Again, isolated algorithms were mentioned in the "Others" field (*e.g.*, GNN, LGBM, CNN, and PCA), altogether summing up less than 15% and not significantly influencing the overall distribution (**P = 14.868 [14.708, 15.028]**).

The most used programming languages in ML-enabled systems projects are presented in Figure 5 (e). Python is by far the most used language (**P = 92.584 [92.459, 92.708]**). For this analysis, we coded and aggregated the programming languages informed in the "Others" field into separate options already shown in the figure.

We also asked participants about the number of projects they have worked on and the number of these that went into production. Analyzing the data, we found that less than half of the projects make it into production (**P = 45.27 [45.11, 45.44]**, with **N = 169**).

### *4.3. RQ2. Perceived Relevance and Complexity of the ML Life Cycle Phases*

The bootstrapped answers revealed that ML practitioners are extremely worried about understanding requirements. The *Problem Understanding and Requirements* phase is clearly perceived as the most relevant and most complex life cycle phase, as presented in Figure 6 (a) and Figure 6 (b), respectively. Nevertheless, it is also possible to observe that, in general, all life cycle phases are perceived as relevant and mainly complex. The exact proportions and bootstrapping confidence intervals were intentionally omitted to improve readability and ease of interpretation. Still, they can be found in our online open science repository [15].

We also asked participants to distribute the percentage of effort among the seven ML life cycle phases (the survey form automatically validated that

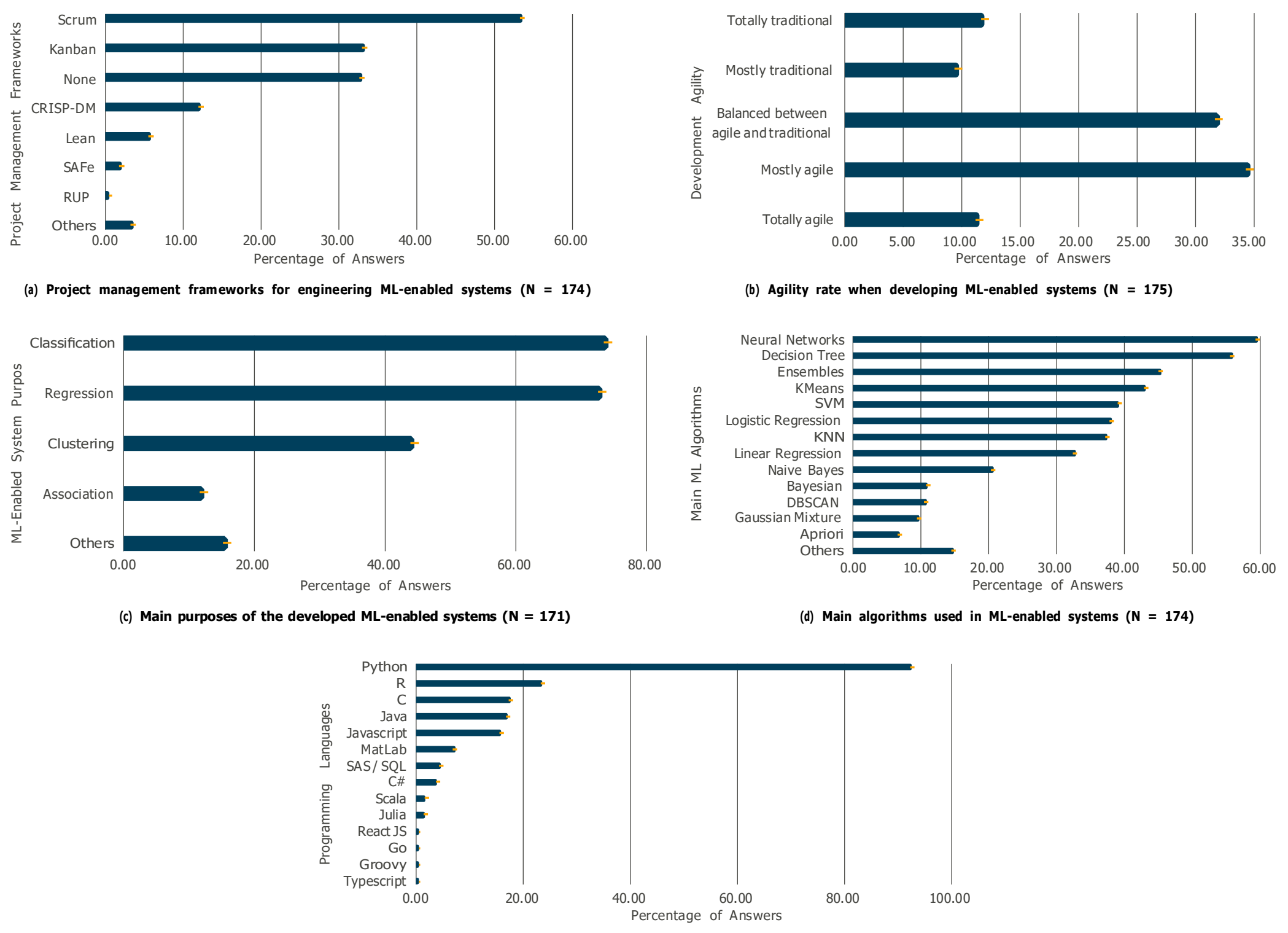

(a) Project management frameworks for engineering ML-enabled systems (N = 174)

(b) Agility rate when developing ML-enabled systems (N = 175)

(c) Main purposes of the developed ML-enabled systems (N = 171)

(d) Main algorithms used in ML-enabled systems (N = 174)

(e) Most used programming languages in ML-enabled systems (N = 176)

Figure 5: ML-enabled systems project context.

the sum of each respondent was equal to 100%). The bootstrapped proportions and confidence intervals are shown in Figure 6 (c). As expected, data pre-processing takes the lead (**P = 18.242 [18.187, 18.297]**), followed by problem understanding (**16.919 [16.876, 16.961]**), data collection (**P = 16.136 [16.093, 16.178]**), and model creation (**P = 15.947 [15.908, 15.986]**).

## *4.4. RQ3. Contemporary RE Practices for ML-enabled Systems*

This question is aimed at revealing how practitioners are currently approaching RE for ML. The answers to each subquestion are presented in Figure 7 and discussed subsequently.

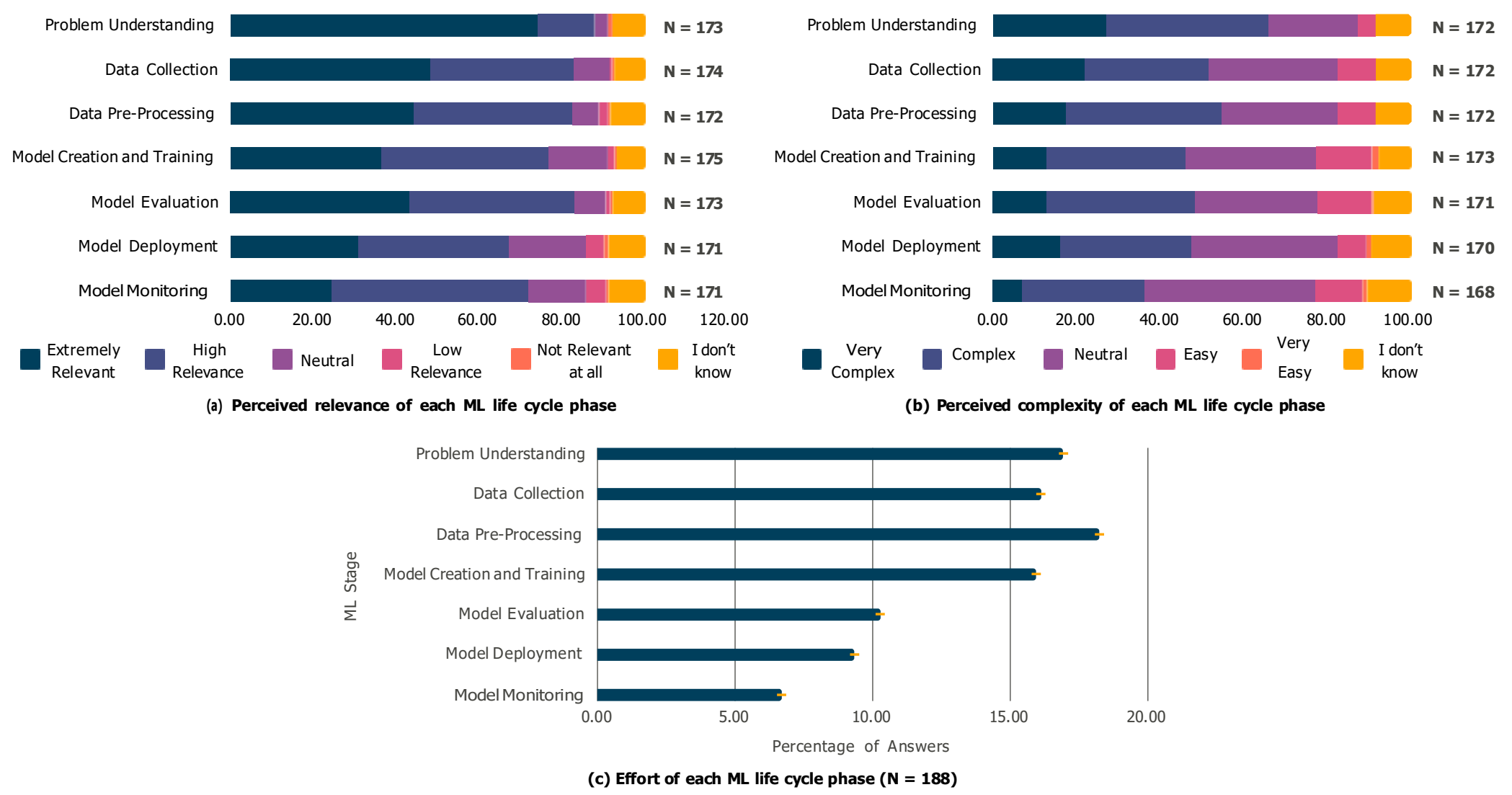

Figure 6: Relevance, Difficulty, and Effort of ML life cycle phases.

#### 4.4.1. *RQ3.1. Who is addressing the requirements of ML-enabled system projects?*

The proportion of roles reported to address the requirements of ML-enabled system projects within the bootstrapped samples and the confidence intervals are shown in Figure 7 (a). It is possible to observe that the project lead (**P = 56.439 [56.17, 56.709]**) and data scientists (**P = 54.71 [54.484, 54.936]**) were most associated with requirements in ML-enabled systems, while business analysts (**P = 29.518 [29.288, 29.749]**) and requirements engineers (**P = 11.202 [11.061, 11.342]**) had a much lower proportion. Several isolated options were mentioned in the "Others" field (*e.g.*, product owner, machine learning engineer, and tech lead), altogether summing up approximately 11% and not significantly influencing the overall distribution (**P = 11.021 [10.865, 11.177]**).

#### 4.4.2. *RQ3.2. How are requirements typically elicited in ML-enabled system projects?*

As presented in Figure 7 (b), respondents reported interviews as the most commonly used technique for requirements elicitation (**P = 55.795 [55.567, 56.022]**), followed (or complemented) by prototyping (**P = 43.953 [43.711, 44.195]**), scenarios **(P = 43.065 [42.834, 43.297]**), workshops (**P = 42.708 [42.483, 42.933]**), and observation (**P = 36.838 [36.613,**

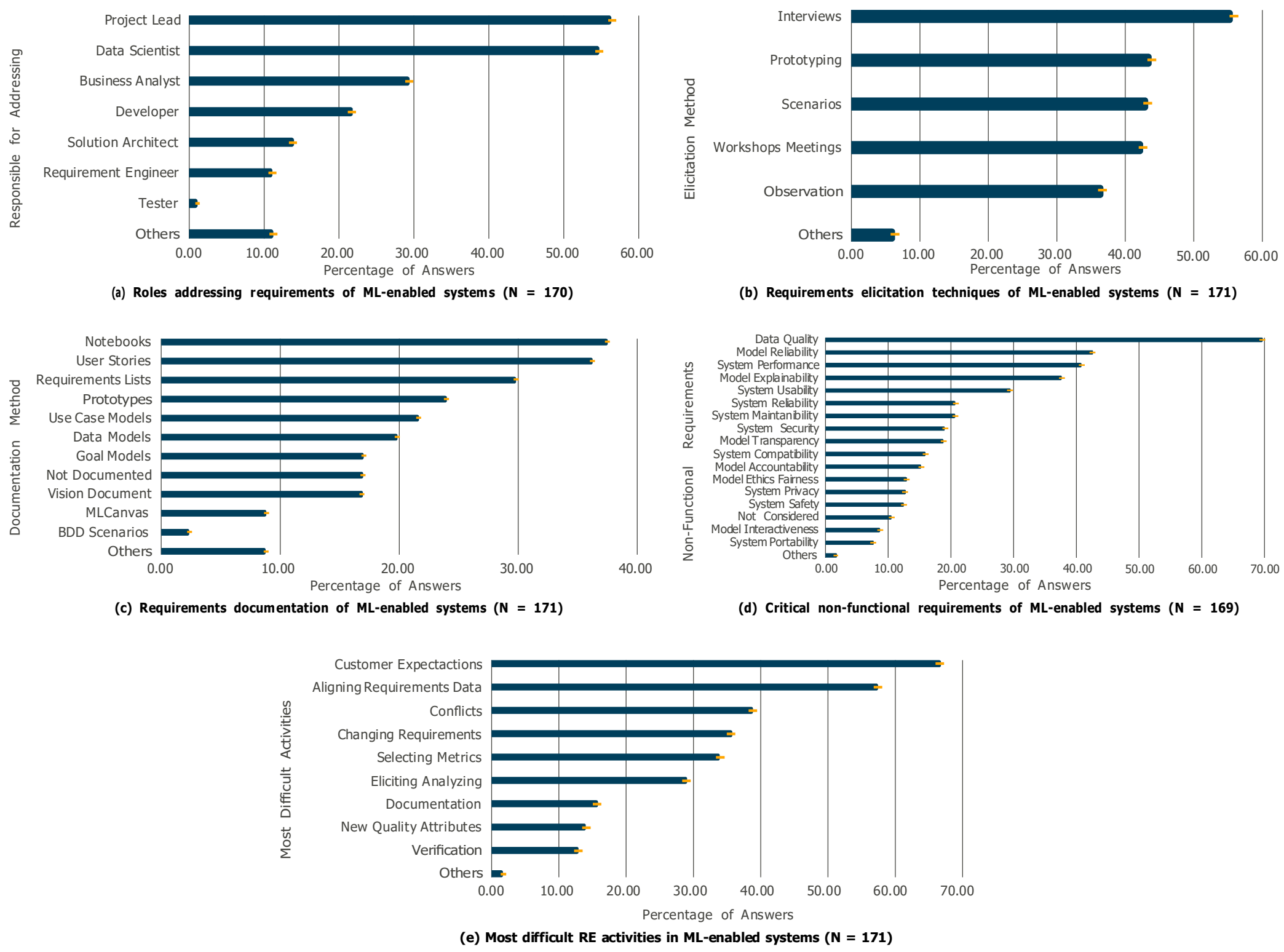

Figure 7: Contemporary RE practices for ML-enabeld systems.

**37.063]**). Some isolated options were mentioned in the "Others" field (**P = 6.488 [6.379, 6.598]**), *e.g.*, meetings with stakeholders, analysis of artifacts, invented, wishful thinking.

#### *4.4.3. RQ3.3. How are requirements typically documented in ML-enabled system projects?*

Figure 7 (c) shows notebooks as the most frequently used documentation format with **P = 37.357 [37.149, 37.564]**, followed by user stories (**P = 36.115 [35.875, 36.356]**), requirements lists (**P = 29.712 [29.499, 29.925]**), prototypes (**P = 23.957 [23.748, 24.166]**), use case models (**P = 21.617 [21.412, 21.822]**), and data models (**P = 19.92 [19.724, 20.117]**). Surprisingly, almost 17% mentioned that requirements are not documented at all (**P = 16.955 [16.767, 17.143]**). Several isolated options were mentioned in the "Others" field (**P = 8.877 [8.744, 9.011]**) *e.g.*, Wiki tools, Google Docs, Jira.

#### *4.4.4. RQ3.4. Which non-functional requirements (NFRs) typically play a major role in ML-enabled system projects?*

Regarding NFRs (Figure 7 (d)), practitioners show a significant concern with some ML-related NFRs, such as data quality (**P = 69.846 [69.616, 70.075]**), model reliability (**P = 42.679 [42.45, 42.907]**), and model explainability (**P = 37.722 [37.493, 37.952]**). Some NFRs regarding the whole system were also considered important, such as system performance (**P = 40.789 [40.573, 41.006]**) and usability (**P = 29.589 [29.36, 29.818]**). A significant amount of participants informed that NFRs were not at all considered within their ML-enabled system projects (**P = 10.617 [10.465, 10.768]**). Furthermore, in the "Others" field (**P = 1.814 [1.745, 1.884]**), a few participants also mentioned that they did not reflect upon NFRs.

#### *4.4.5. RQ3.5. Which activities are considered most difficult when defining requirements for ML-enabled systems?*

We provided answer options based on the literature on requirements [34] and RE for ML [31, 32], leaving the "Other" option to allow new activities to be added. As shown in Figure 7 (e), respondents considered that managing customer expectations is the most difficult task (**P = 66.804 [66.575, 67.032]**), followed by aligning requirements with data (**P = 57.306 [57.066, 57.546]**), resolving conflicts (**P = 38.582 [38.341, 38.824]**), managing changing requirements (**P = 35.62 [35.395, 35.846]**), selecting metrics (**P = 33.95 [33.723, 34.176]**), and elicitation and analysis (**P = 29.036 [28.824, 29.248]**).

### *4.5. RQ4. Contemporary Model Deployment and Monitoring Practices*

This question is aimed at revealing the practices related to the model deployment and monitoring phases. The answers to each subquestion are presented in Figure 8 and discussed subsequently.

#### *4.5.1. RQ4.1. Which approaches are used to deploy ML models?*

Participants were asked about which approach they usually take for hosting their models, where respondents could select more than one option. As shown in Figure 8 (a), deploying models as a separate service was the most frequent choice (**P = 59.457 [59.219, 59.695]**), followed by embedding models into the consuming application (**P = 42.719 [42.476, 42.962]**) and PaaS ML solutions (**P = 23.826 [23.628, 24.024]**). Other reported options (**P = 5.47 [5.359, 5.58]**) that were mentioned included using specific containers and technical reports.

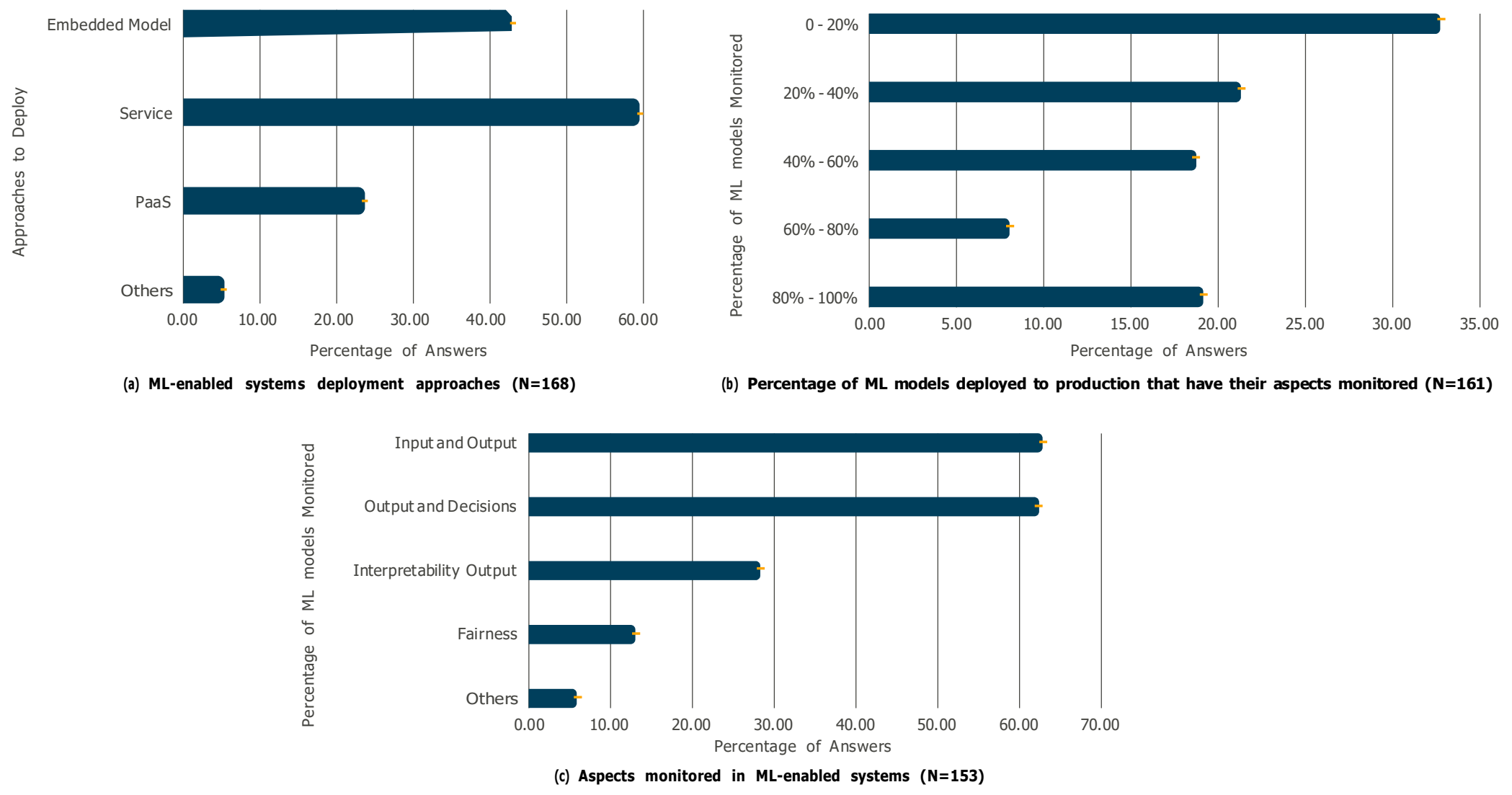

Figure 8: Contemporary Model Deployment and Monitoring practices for ML-enabeld systems.

*4.5.2. RQ4.2 What are the MLOps practices and principles used?*

We asked if the respondents’ organizations follow MLOps practices or principles. We received 168 answers for this question (N = 168). The majority (a bootstrapping proportion of more than 70%) answered not having MLOps practices or principles in place at their organizations (**P = 70.926 [70.703, 71.149]**). Those who answered having MLOps practices or principles in place (**P = 29.074 [28.851, 29.297]**) had the option to provide details on their MLOps approach in open text format.

Of the 49 respondents with MLOps practices or principles in place, 23 provided textual details. These answers were diverse and ranged from conducting regular retraining and deployment with data snapshots to having MLOps pipelines built on top of a continuous delivery tools (*e.g.*, Gitlab CI/CD, Azure DevOps, Kubernetes), and using specific ML development platforms to manage ML pipelines (*e.g.*, Apache Airflow, AWS Sagemaker MLOps, Azure ML Studio, BentoML, Databricks, Dataiku, MLFlow, Vertex AI). These platforms typically enable the implementation of MLOps prac-tices such as drift detection, model re-training, and monitoring of relevant aspects. Some of these answers also reported more informally that they were using cloud-based solutions and that their MLOps efforts were in progress or being implemented.

#### *4.5.3. RQ4.3. What percentage of projects deployed into production have their ML models being monitored?*

We asked participants explicitly about the percentage of projects deployed into production that have their ML models being monitored. The bootstrapping results indicate that less than 40% of the models in production are being monitored **(P = 38.69 [38.50, 38.83]**).

We analyzed the percentage of participants within different percentage ranges of models being monitored to better understand the distribution. Figure 8 (b) summarizes this analysis. It shows that the most frequent range is the one with less than 20% of the models being monitored (**P = 32.729 [32.505, 32.952]**), followed by the range from 20% to 40% (**P = 21.309 [21.118, 21.5]**). While this is alarming, given that ML-enabled systems are being used in critical business sectors (*cf.* Figure 4), still a considerable amount of almost 20% of the participants reported that 80% to 100% of their models are being monitored (**P = 19.153 [18.961, 19.344]**).

#### *4.5.4. RQ4.4. Which aspects of the models are monitored?*

Respondents described which aspects (multiple selections allowed) were actually monitored as in Figure 8 (c). Inputs and outputs are the most frequently monitored aspect (**P = 62.832 [62.582, 63.082]**), followed closely by outputs and decisions (**P = 62.405 [62.164, 62.645]**). Some ML-enabled system projects also monitor interpretability output (**P = 28.32 [28.1, 28.539]**). Fairness is less frequently monitored (**P = 13.046 [12.878, 13.214]**). The "Others" field was filled by a minority (**P = 5.876 [5.758, 5.993]**) and included respondents reporting to monitor feature drifting and performance (CPU, memory) logs.

### *4.6. RQ5. Main Problems of ML-enabled System Projects*

This question aimed to identify the main problems for each ML life cycle phase and the ones most likely leading to overall project failure. We applied open and axial coding procedures to allow the problems to emerge from open-text responses provided by the practitioners. First, participants could inform up to three problems related to each ML life cycle phase. Thereafter, they were asked to inform the top three problems leading to overall project failure.

After the open coding of the responses, we incorporated axial coding procedures to provide an easily understandable overview, relating the emerging codes to categories. We started with the categories *Input*, *Method*, *Organization*, *People*, and *Tools*, as suggested for problems in previous work on defect causal analysis [11]. Based on the data, we renamed the *Tools* category into the more generic *Infrastructure* category. Furthermore, as problems concerned ML-enabled systems that basically learn from data, we identified the need to add a new category for problems related to *Data*. These categories were identified considering the overall coding efforts for the problems related to the seven ML life cycle phases and the problems leading to project failure.

To ease understanding, we graphically illustrate the problems by provid- ing an overview of the frequencies of the resulting codes using a probabilistic cause-effect diagram, which was introduced for causal analysis purposes in our previous work [13, 14]. While this diagram provides a comprehensive overview, we emphasize that, in this context, the percentages do not repre- sent probabilities but just frequencies of occurrence of the codes (*i.e.*, the sum of all code frequencies is 100%). This diagram also provides the sum for the frequencies of problems within each individual category and organizes the problems with the highest frequencies within each category closer to the middle. These diagrams have been successfully used for a similar purpose in the Naming the Pain in Requirements Engineering research project [5].

#### *4.6.1. RQ5.1. What are the main problems reported for each ML life cycle phase?*

Participants could inform up to three problems related to each ML life cycle phase. We provide the coded and categorized results for each ML life cycle phase hereafter. It is noteworthy that these codes were produced by one researcher (the fifth author), peer-reviewed by another researcher (the first author), and then had a double independent validation (by the second and the third author). Furthermore, the original texts, codes, and categories are available for auditing in our online open science repository [15].

*Problem Understanding and Requirements.* Considering only the answers of participants who completed all parts of the survey, we had 249 open-text answers on problems analyzed for the problem understanding and requirements ML life cycle phase. The results are shown in Figure 9 (a).

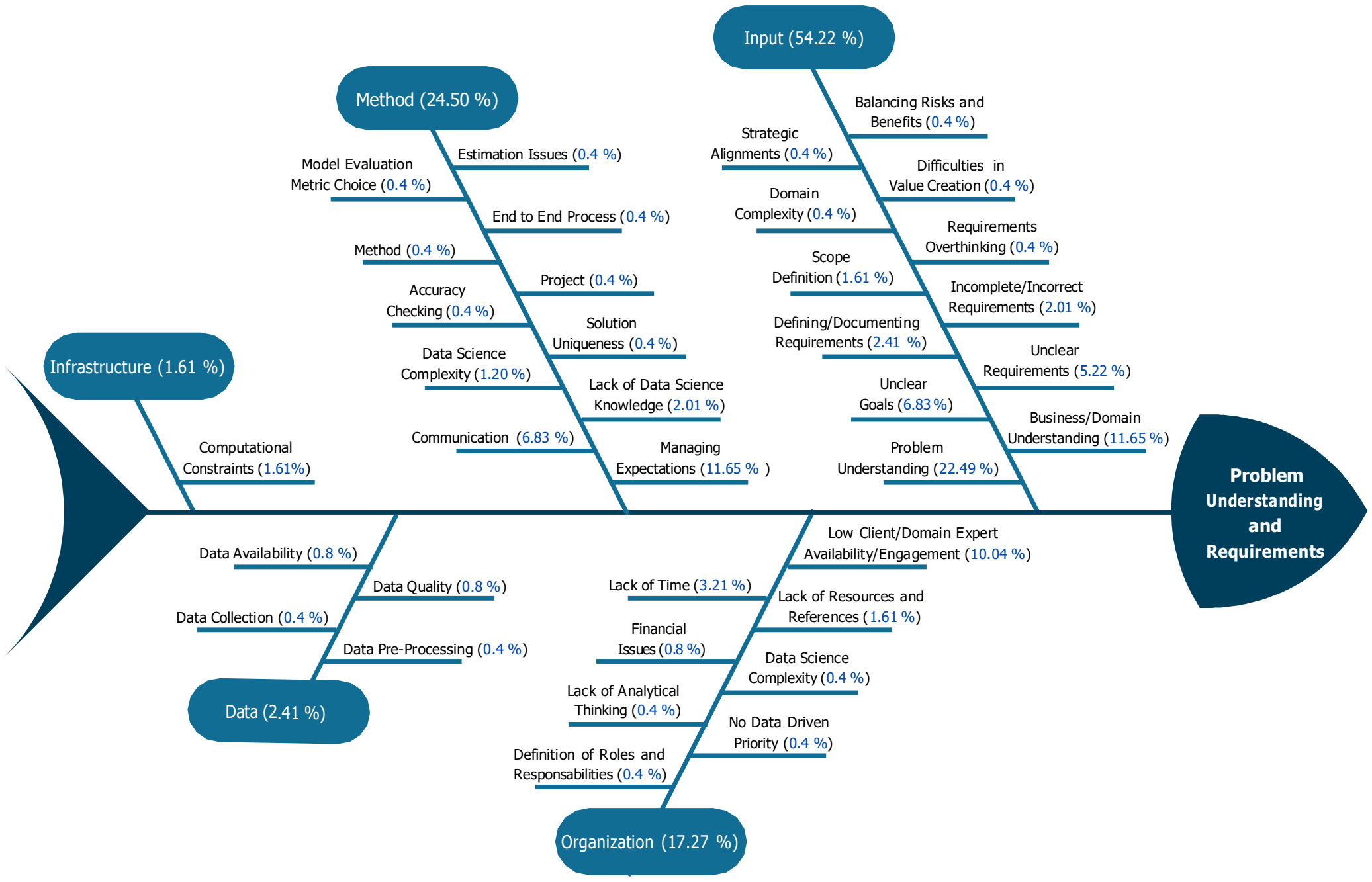

**(a) Problems related to the Problem Understanding and Requirements ML life cycle phase**

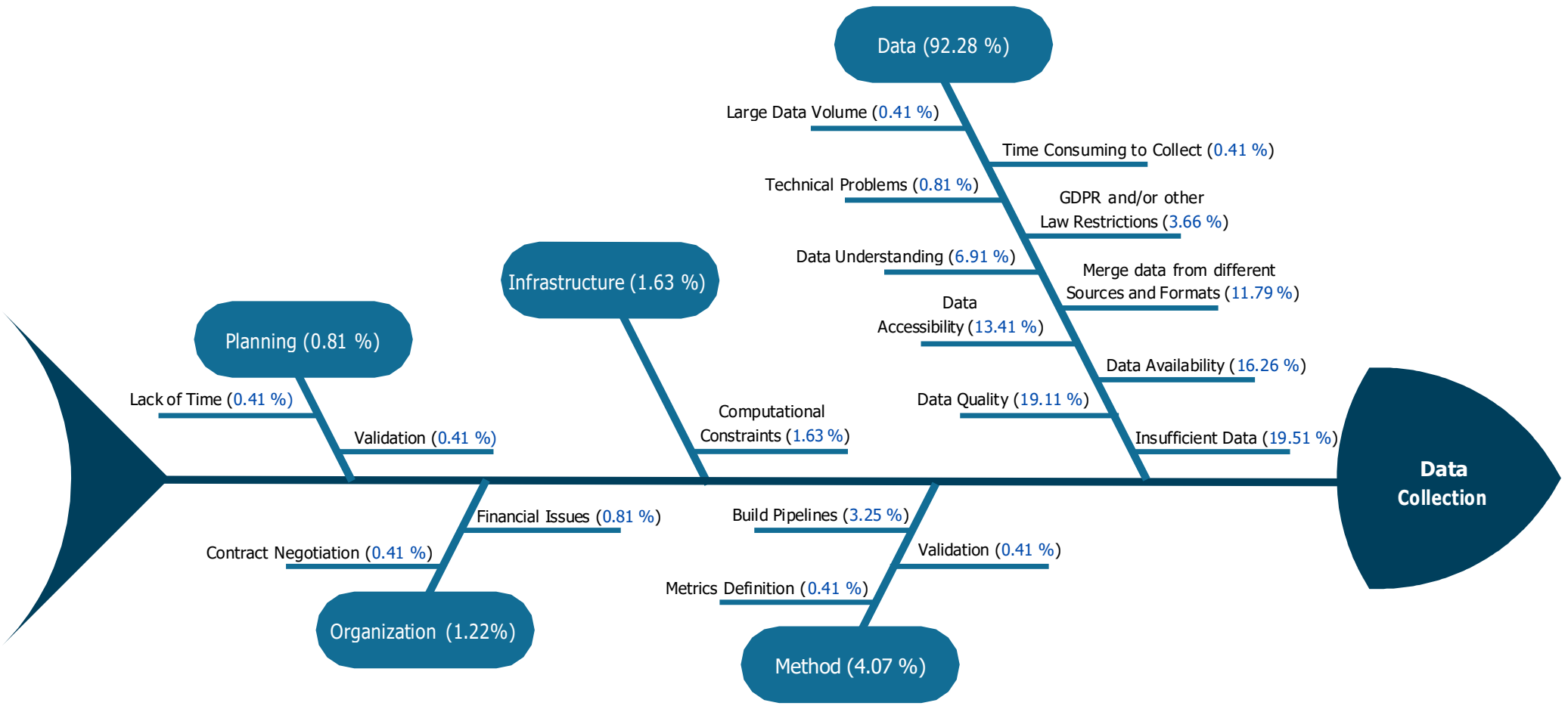

**(b) Problems related to the Data Collection ML life cycle phase**

Figure 9: Problems related to the Problem Understanding and Requirements and Data Collection ML life cycle phases.

It is possible to observe that most of the reported problems are related to the *Input* category, followed by *Method* and *Organization*. Within the *Input* category, the main reported problems concern difficulties in understanding the problem and the business domain and unclear goals and requirements. In the *Method* category, the prevailing reported problems concern difficulties in managing expectations and establishing effective communication. Finally, in the *Organization* category, the lack of customer or domain expert availability and engagement and the lack of time dedicated to requirements-related activities were mentioned. While we focus our summary on the most frequently mentioned problems, it is noteworthy that the less frequent ones may still be relevant in practice. For instance, computational constraints or a lack of data quality (or availability) can directly affect ML-related possibilities and requirements.

*Data Collection.* Using the same criteria of considering only answers from participants who completed all parts of the survey, we had 246 open-text answers on problems analyzed for the data collection ML life cycle phase. The results are shown in Figure 9 (b). As this ML life cycle phase is closely related to data, it was expected to have most of the reported problems related to the *Data* category. Within the *Data* category, the main reported problems concern insufficient data and data quality and availability issues. Data accessibility, data understanding, and difficulties in merging data from different sources and formats were also mentioned relatively often. Again, we emphasize that the less frequent prob- lems may still be relevant in practice. For instance, depending on the con- text, GDPR and law restrictions related to privacy may become particularly relevant issues to be addressed for data collection.

*Data Pre-processing.* Participants who completed all parts of the survey provided 244 open-text answers on problems for the data pre-processing ML life cycle phase. The coded results are shown in Figure 10 (a). As this ML life cycle phase is also closely related to data, most of the reported problems were related to the *Data* category. Within the *Data* cat- egory, the main reported problems related to pre-processing concern issues related to data quality, merging data from different sources and formats, and data cleaning. Specific data pre-processing challenges were also mentioned, such as understanding the data and dealing with insufficient data *(e.g.*, with data augmentation techniques) and missing values. As examples from other

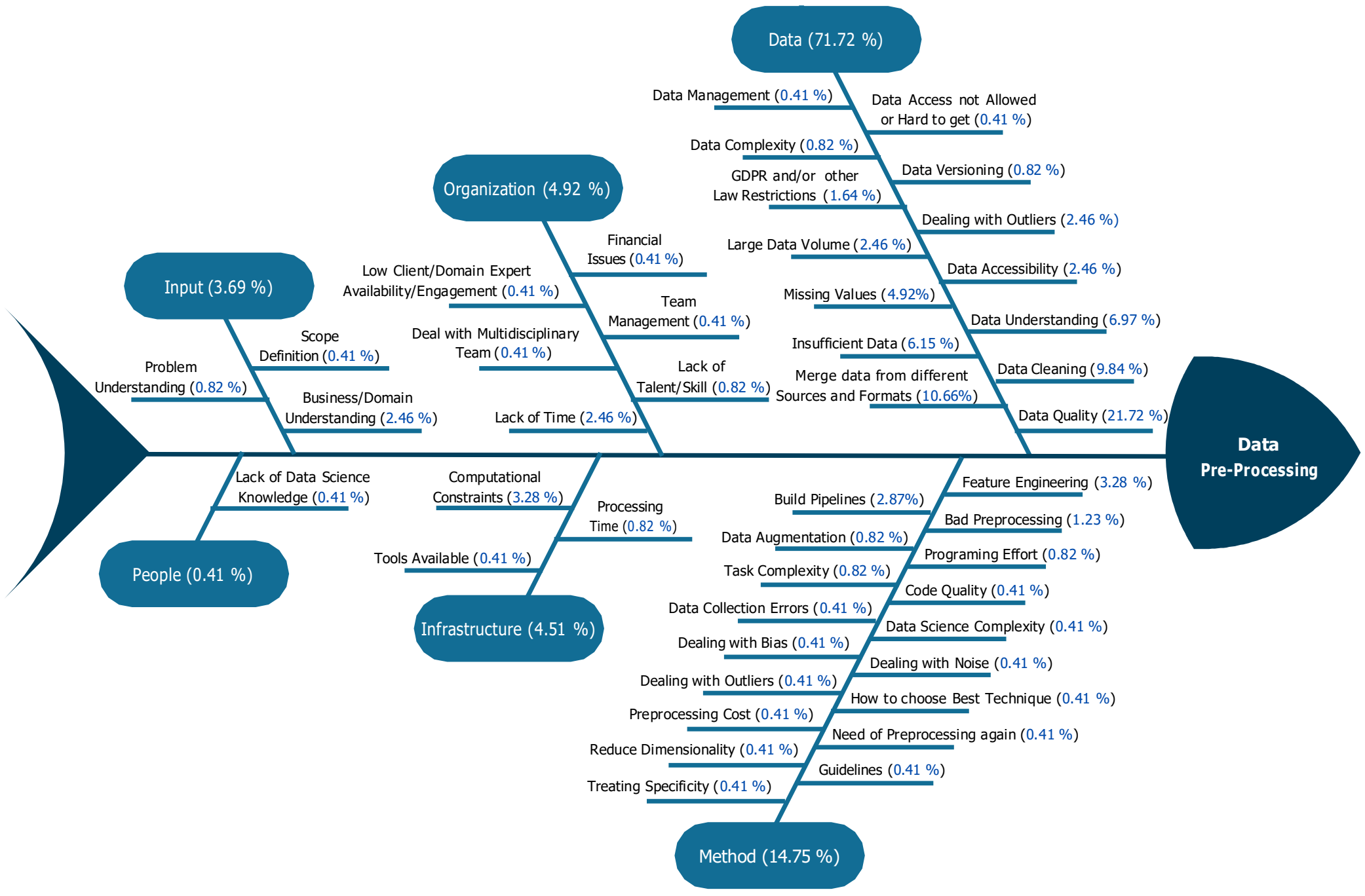

**(a) Problems related to the Data Pre-Processing ML life cycle phase**

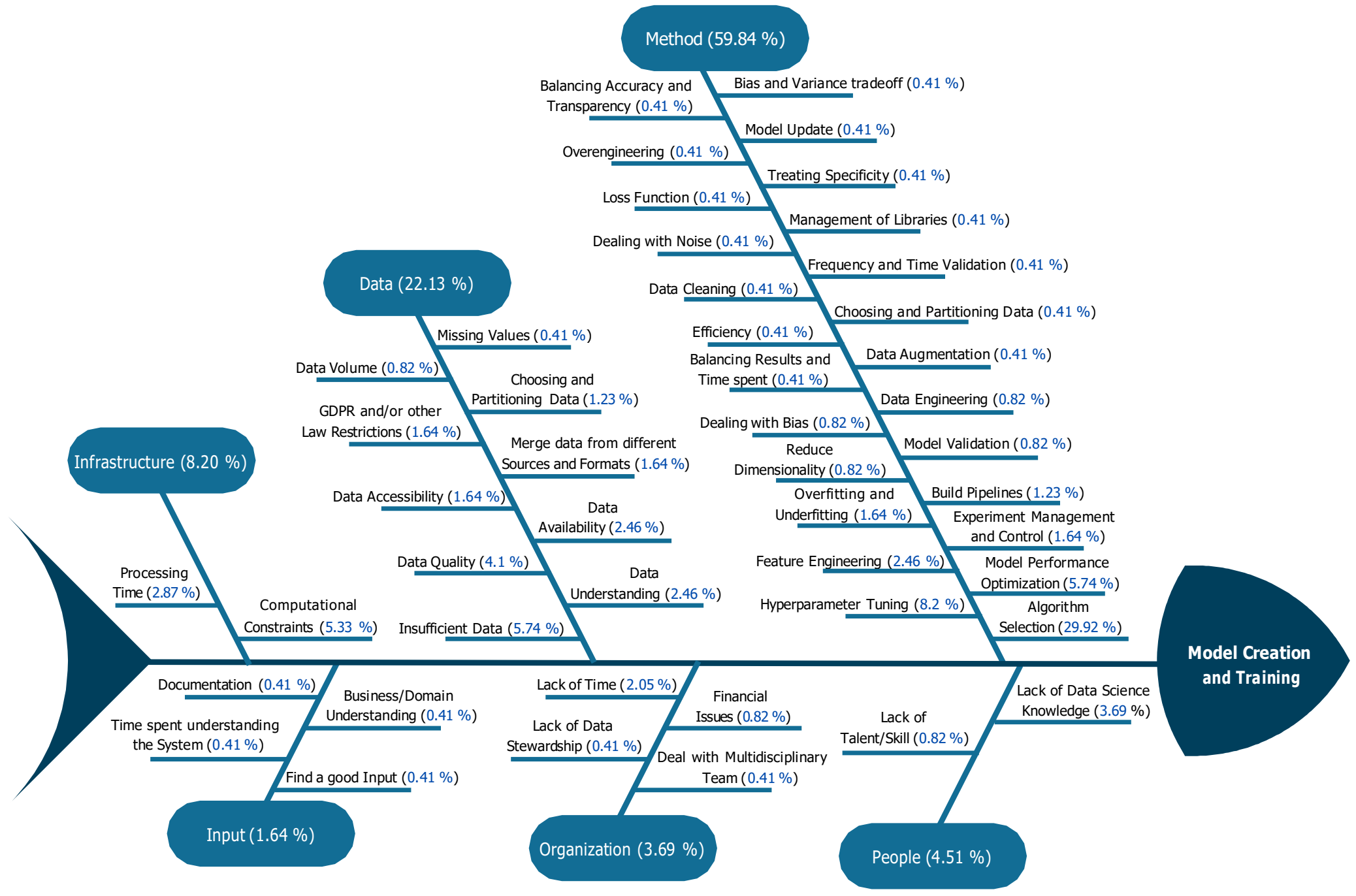

**(b) Problems related to the Model Creation and Training ML life cycle phase**

Figure 10: Problems related to the Data Pre-Processing and Model Creation and Training ML life cycle phases.

categories, in the *Method* category, feature engineering (*i.e.*, selecting the features to build the model) was mentioned. In the *Infrastructure* category, computational constraints for pre-processing were also mentioned as a problem by some of the participants.

*Model Creation and Training.* Following our selection criteria, analyzing problems mentioned for model creation and training involved coding 244 open-text answers. The coded results are shown in Figure 10 (b). As expected, for this ML life cycle phase, most of the reported prob- lems were related to the *Method* category. Within the *Method* category, the most frequently reported problems concerned selecting the algorithm, hyper- parameter tuning, and model performance optimization. Within the other categories, insufficient data, computational constraints to train the model, and lack of data science knowledge were also frequently mentioned.

*Model Evaluation.* Analyzing problems mentioned for model evaluation involved coding 157 open-text answers. The coded results are shown in Figure 11 (a). As expected, for the model evaluation life cycle phase, most of the re- ported problems were also related to the *Method* category. The most fre- quently reported problems concerned selecting the best metrics and tech- niques for model evaluation, correctly interpreting the results, and han- dling model performance issues (*e.g.*, underfitting, overfitting). Building the dataset was also frequently mentioned, referring to difficulties in developinga reliable dataset for model evaluation.

*Model Deployment.* We had 138 open-text answers for problems related to model deployment. The coded results are shown in Figure 11 (b). As per the survey respondents, the top problems faced within the de- ployment phase were related to preparing the infrastructure for production deployment, the difficulty of integrating legacy applications, defining what architectural infrastructure to use, how to scale it, and financial issues.

*Model Monitoring.* Finally, we had 112 open-text answers for problems related to model monitoring. Figure 12 (a) presents the results of the open and axial coding of the answers for the main problems of the monitoring phase. Here, the most frequently reported problems by the participants were related to the need to develop their own monitoring tools and to evaluate and choose the appropriate metrics while mainly not having experience in monitoring models and building monitoring platforms.

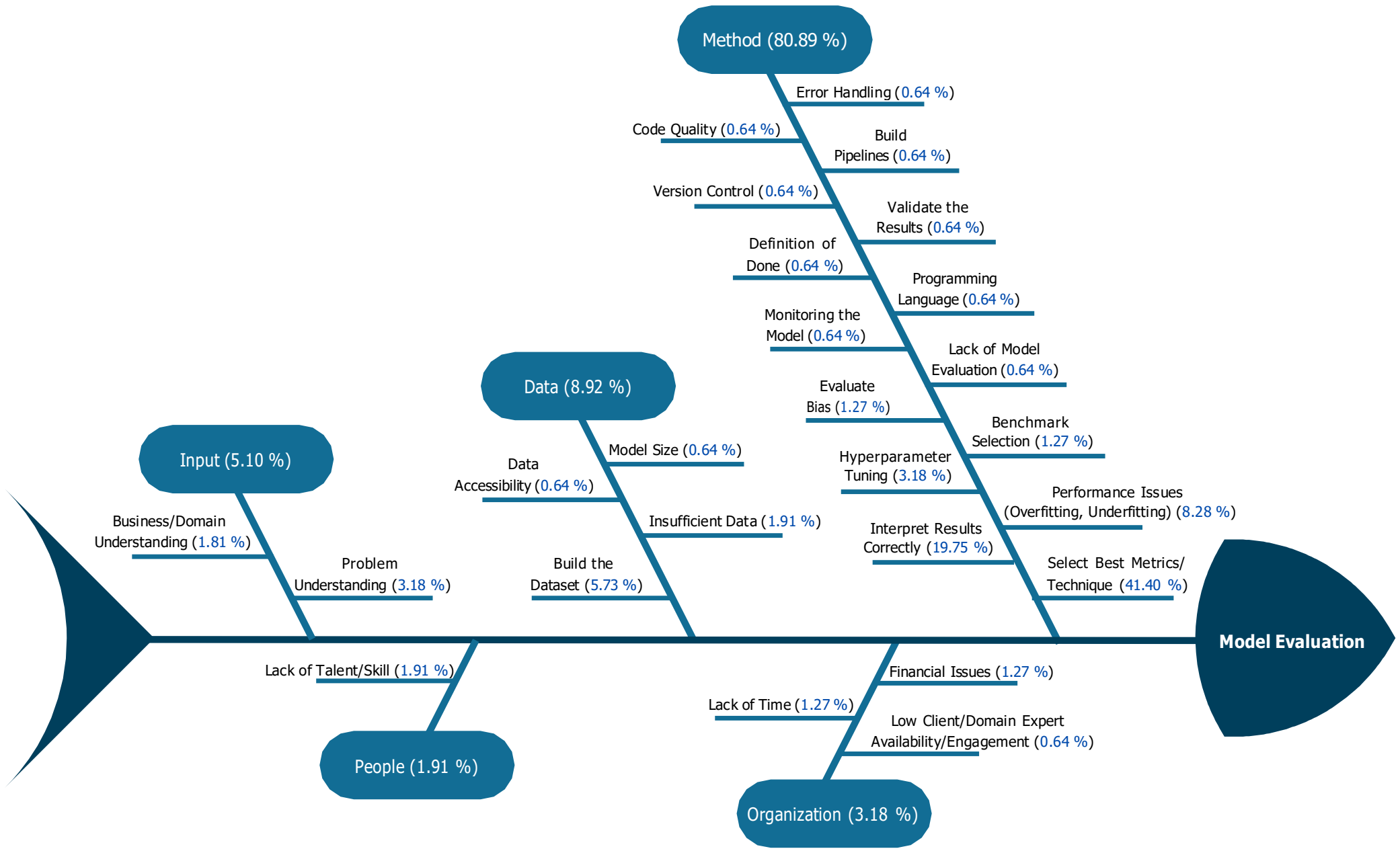

**(a) Problems related to the Model Evaluation ML life cycle phase**

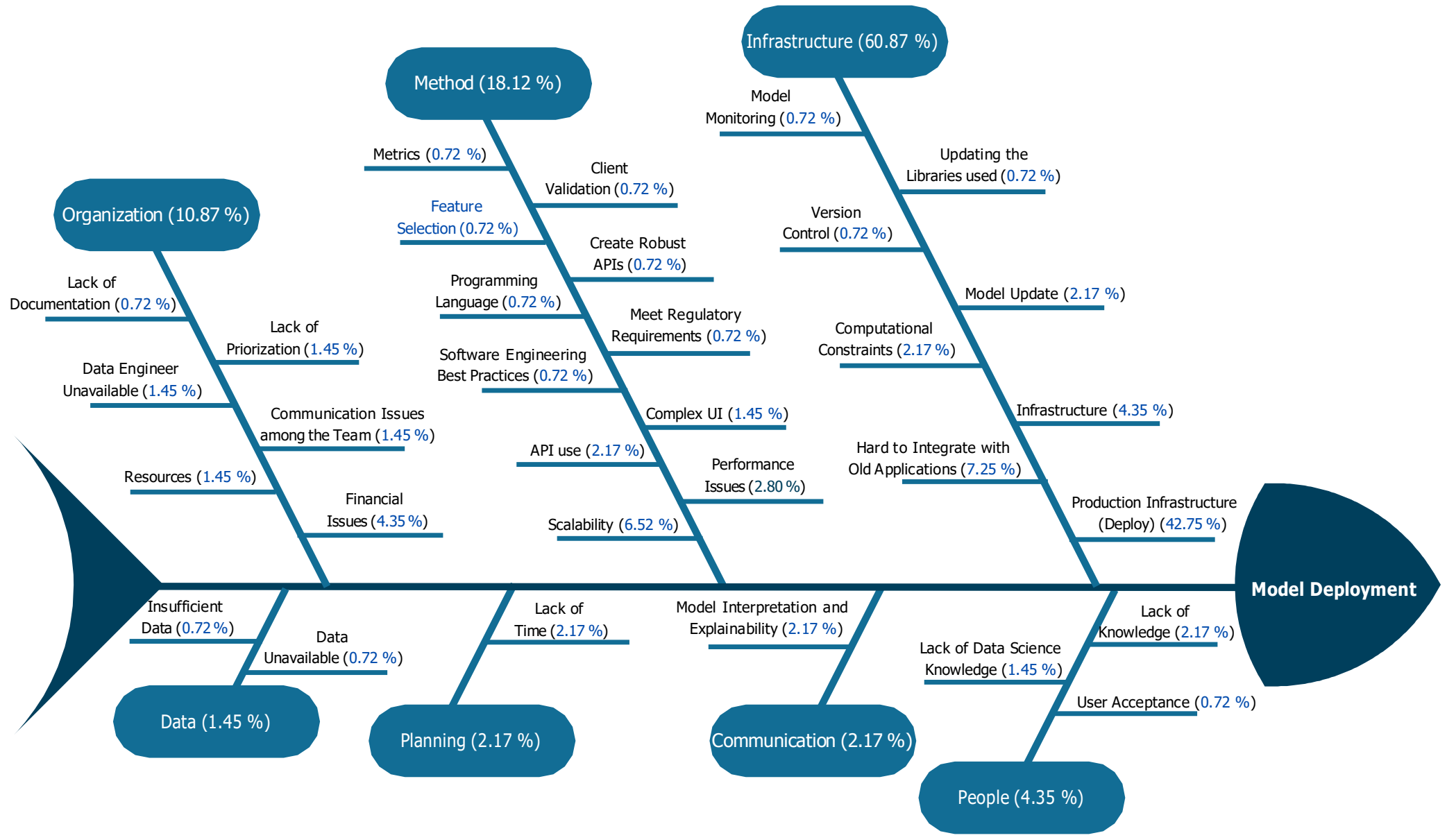

**(b) Problems related to the Model Deployment ML life cycle phase**

Figure 11: Problems related to the Model Evaluation and Model Deployment ML life cycle phases.

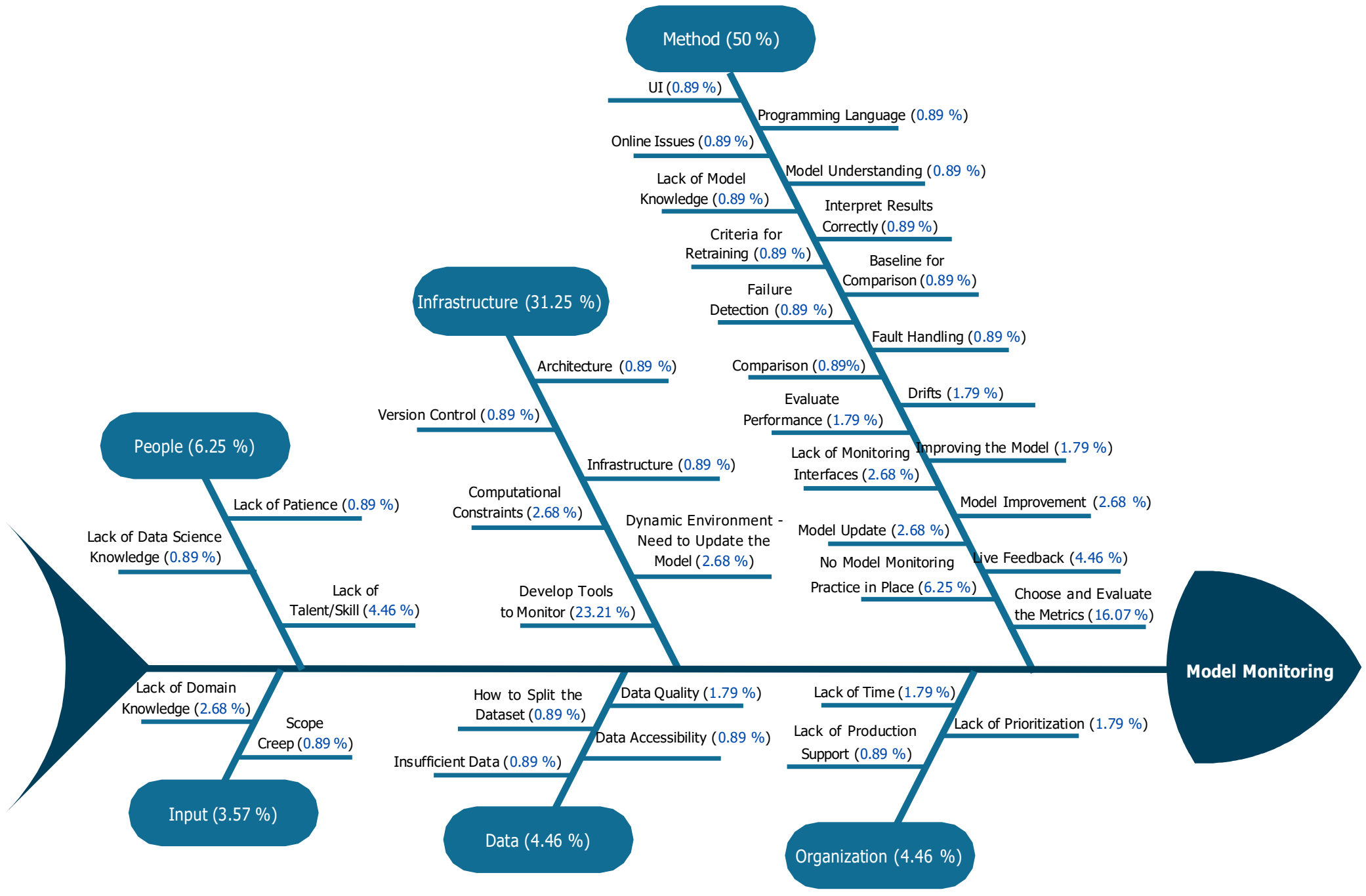

**(a) Problems related to the Model Monitoring ML life cycle phase**

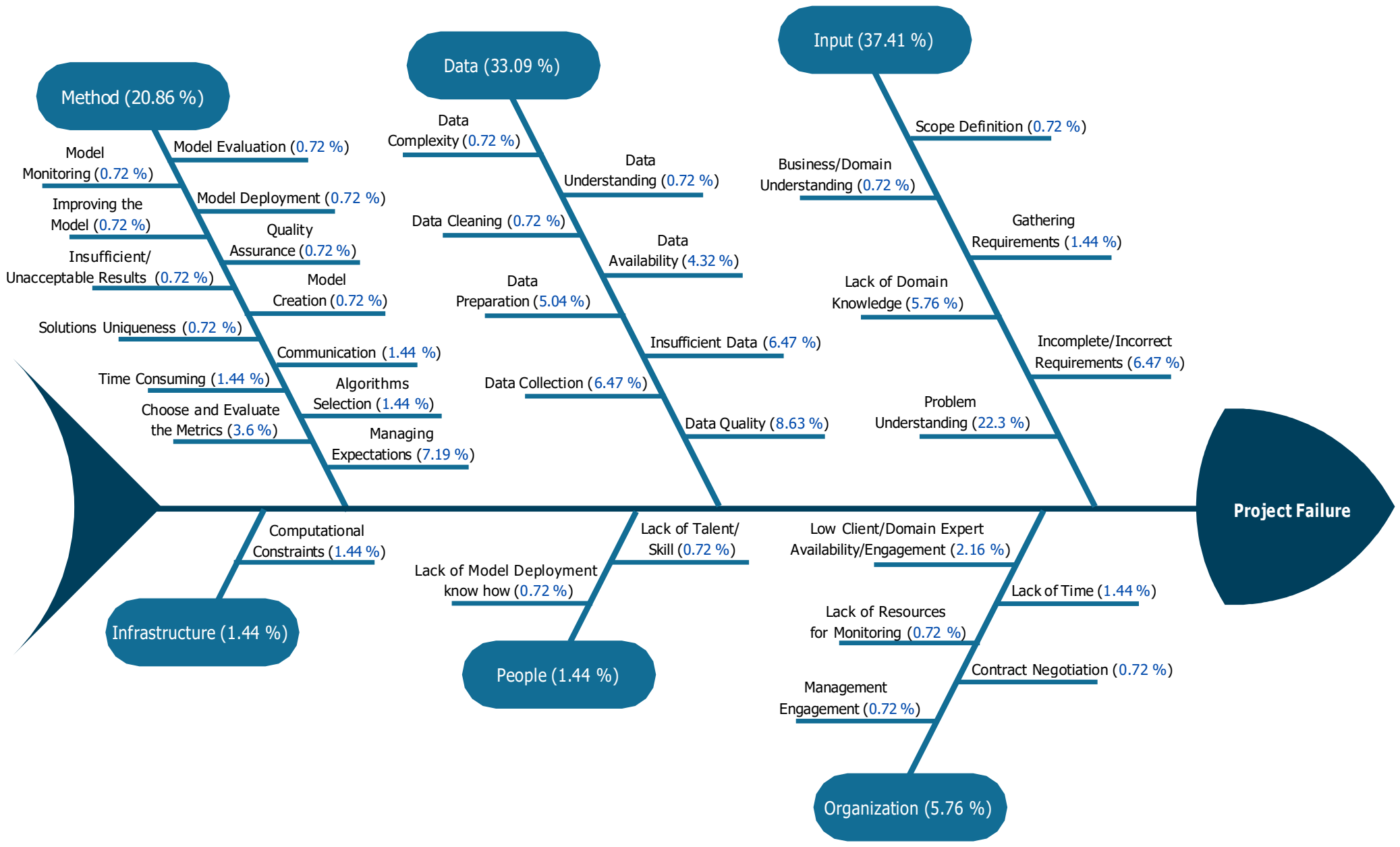

**(b) Problems that lead to Project Failure**

Figure 12: Problems related to the Model Monitoring ML life cycle phase and problems that lead to project failure.

#### *4.6.2. RQ5.2. What problems were reported as leading to overall ML-enabled system project failure?*

Participants were asked to inform the top three problems leading to overall project failure. In this context, project failure was defined as a project that did not result in a released product or was canceled due to not providing the expected results. As we were analyzing project failure, we conservatively analyzed only the first (most critical) problem reported by the participants. The second and third most critical problems reported were discarded from the analysis. Participants informed 139 open-text answers for the most critical problem leading to project failure. The results of the open and axial coding of the answers are shown in Figure 12 (b).

It is possible to observe a wide range of problems spanning through different categories, with *Input*, *Data*, and *Method* being the most frequent categories. Aligned with our previous findings on the relevance of the ML life cycle phases, difficulties related to problem understanding were the most frequently reported problem in the open-text answers, followed by issues related to data quality and to managing expectations. These results highlight the important role of requirements and data in preventing ML-enabled system project failure.

## 5. Discussion

In this section, we discuss our survey's results and main findings. Based on the research questions, we group the discussion into ML-enabled systems project context, relevance and complexity of ML life cycle phases, contemporary practices of RE for ML, contemporary practices for deploying and monitoring ML models, and main reported problems.

*ML-Enabled Systems Project Context.* Our results reveal a preference for agile methodologies, particularly Scrum, used by over half of the respondents. These findings align well with what is observed in general software engineering contexts [17] and with the variable scope and experimental nature of work required in ML projects. However, we also observed that many projects do not use any project management framework, suggesting a possible need to further adapt agile management frameworks for this context. Indeed, the community is already taking some efforts in this direction (*e.g.*, [30]).

Continuing within the context of ML-enabled system projects, classification and regression are the types of problems predominantly tackled. This focus reflects their foundational role in ML applications across various sectors.

With respect to algorithms, we observed extensive use of neural networks, decision trees, and ensembles combining different models. Additionally, Python stands out as the most used programming language, utilized by more than 90% of the participants.

An alarming finding is that less than half of the ML-enabled system projects make it into production. While close, the scenario indicated by our result is even worse than the 53% reported by Gartner [6].

*Perceived Relevance and Complexity of ML Life Cycle Phases.* The results highlight the critical importance of the problem understanding and requirements phase, perceived as the most relevant and complex part of the ML life cycle. This finding is aligned with the literature, suggesting that defining requirements reflecting what the ML-enabled system is expected to achieve is challenging [33, 31, 32], likely due to the abstract nature of translating business problems into ML tasks. We also observed that significant effort is required in these projects for data collection and, particularly, data preprocessing, reflecting the data-centric nature of ML projects.

*Contemporary RE Practices for ML-Enabled Systems.* We found significant differences in RE practices within ML-enabled system projects compared to RE practices in conventional SE projects [5]. For instance, in ML-enabled system projects, RE-related activities are mostly conducted by project leaders and data scientists, and the prevalent requirements documentation format concerns interactive Notebooks. The main focus of NFRs in this context becomes data quality, model reliability, and model explainability. Furthermore, the identified difficulties in managing customer expectations and aligning requirements with data highlight the challenge also reported in the literature of bridging the gap between business objectives and technical capabilities in ML projects [33, 31, 25]. This scenario indicates the need to adapt and disseminate RE-related practices for engineering ML-enabled systems. The community is already taking efforts in this direction. For requirements specification, for instance, Villamizar *et al.* [32] proposed a specific approach for specifying ML-enabled systems that showed being effective in industrial evaluations [32].

*Model Deployment and Monitoring Practices.* With respect to model deployment, our results reveal that models are mainly deployed as separate services and that embedding the model within the consuming application or platform-as-a-service solutions are less frequently explored. Most practitioners do not follow MLOps principles and do not have an automated pipeline to retrain and redeploy the models. Indeed, literature reports that building end-to-end ML pipelines is challenging due to the difficulties of integrating ML and non-ML components, the complexity of integrating many tools, and the need for engineering skills beyond the typical comfort zone of data scientists [25, 27]. Regarding monitoring, the findings that less than 40% of ML models are monitored post-deployment is concerning. This lack of monitoring can lead to undetected issues, affecting the reliability and performance of ML-enabled systems. Participants mainly reported not having well-established monitoring practices at their organizations and difficulties choosing appropriate metrics. Difficulties aligned with the ones revealed in our survey have been reported. Nahar *et al.* [25], for instance, found that even for companies that adopt a monitoring infrastructure, practitioners report struggling with ad-hoc monitoring practices. Furthermore, our results reveal that the main monitored aspects in ML-enabled system projects are inputs, outputs, and decisions taken. Other relevant aspects, such as model fairness, are typically not being monitored.

*Main Problems of ML-Enabled System Projects.* Identifying prevalent problems across the different life cycle phases provides insights into where organizations struggle when engineering ML-enabled systems. We applied qualitative analysis procedures to allow the problems to emerge from open-text responses. The diagrams (*cf.* Figures 9 to 12) depict the problems that reflect the intricacies of each ML life cycle stage, naming the pain based on the conducted qualitative analyses. We emphasize the most critical problems leading to project failure. Difficulties with problem understanding was by far the most frequently reported problem, followed by issues related to managing expectations and data quality. The frequent mention of problems related to understanding project requirements and managing expectations reflects the ongoing challenges in effectively communicating and setting realistic project goals. On the other hand, the frequent mention of data quality issues highlights the important role that high-quality data plays in the success of ML projects.

Overall, our study complements other valuable research on challenges related to ML-enabled systems (*e.g.*, the systematic review on the state-of-the-art by Giray [7], the extensive literature meta-summary by Nahar *et al.* [25]), strengthening their findings with results from a large-scale international survey, and providing complementary details on contemporary practices and a detailed overview of the main problems as reported by practitioners.

## 6. Threats to Validity

We identified some threats while planning, conducting, and analyzing the survey results. Hereafter we list these potential threats, organized by the survey validity types presented in [20].

**Face and Content Validity**. Face and content validity threats include bad instrumentation and inadequate explanation of constructs. To mitigate these threats, we involved several researchers in reviewing and evaluating the questionnaire with respect to the format and formulation of the questions, piloting it with 18 Ph.D. students for face validity and with five experienced data scientists for content validity.

**Criterion Validity**. Threats to criterion validity include not surveying the target population. We clarified the target population in the consent form (before starting the survey). We also considered only complete answers (*i.e.*, answers of participants that answered all four survey sections) and excluded participants that informed having no experience with ML-enabled system projects.

**Construct Validity**. We ground our survey questions and answer options on theoretical background from some of our previous studies investigating contemporary problems [5, 34] and on typical ML life cycle phases [2, 12]. Furthermore, to gather answer options for contemporary practices, we analyzed the literature on RE for ML [33, 31, 32] and model deployment and monitoring [10, 9, 27]. Still, all questions with predefined answer options also include the "Others" option in open-text format. A threat to construct validity is inadequate measurement procedures and unreliable results. To mitigate this threat, we follow recommended data collection and analysis procedures [35], employing bootstrapping with confidence intervals for quantitative analyses and open and axial coding procedures for qualitative analyses.

**Reliability**. One aspect of reliability is statistical generalizability. We could not construct a random sample systematically covering different types of professionals involved in developing ML-enabled systems, and there is yet no generalized knowledge about what such a population looks like. Furthermore, as a consequence of convenience sampling, the majority of answers came from Europe and South America. Nevertheless, the experience and background profiles of the subjects are comparable to the profiles of ML teams as shown in Microsoft's study [16]. To deal with the random sampling limitation, we used bootstrapping and conservatively reported confidence intervals, avoiding null hypothesis testing. Another reliability aspect concerns inter-observer reliability, which we improved by including two independent peer reviews in all our qualitative analysis procedures and making all the data and analyses openly available online [15].

## 7. Concluding Remarks

This study provides a comprehensive and empirically grounded overview of the current practices and challenges in engineering ML-enabled systems. By conducting an international survey with 188 practitioners from 25 countries, we have not only affirmed the findings of previous research but also expanded on them, offering additional insights into the status quo and into problems related to real-world complexities faced by professionals in the field.

Our findings reveal a preference for agile methodologies, particularly Scrum, in managing ML-enabled system projects, aligning with ML development’s adaptive and iterative nature. However, a notable proportion of projects lacking management practices suggests a gap that could potentially be addressed by tailoring agile methodologies more closely to the demands of ML projects.

Regarding the ML life cycle, the problem understanding and requirements phase stands out as the most relevant and the most complex, highlighting the ongoing challenge of translating business problems into actionable ML tasks. We also observed significant efforts devoted to data collection and data preprocessing, which can be explained by the data-centric nature of ML projects.

Our study reveals that RE practices for ML systems notably diverge from traditional software projects, *e.g.*, with project leaders and data scientists often leading RE activities and notebooks being used as the primary documentation format. This scenario and the identified challenges in aligning business objectives with technical capabilities and managing customer expectations call for adaptations of RE practices to better support ML-enabled systems engineering. Additionally, we identify critical gaps in model deployment and monitoring, with many ML-enabled system projects lacking automated pipelines for retraining and redeploying models, and mainly missing or inadequate post-deployment monitoring practices.

The qualitative analysis of problems reported by the practitioners provides a detailed map of the challenges faced within each ML life cycle phase.

We believe that the mapping of these challenges can help steer research on engineering ML-enabled systems in a problem-driven manner. Furthermore, it is noteworthy that some of these challenges (*e.g.*, issues with problem understanding, expectation management, and data quality) are perceived as particularly critical and leading to overall project failure.

We believe that the research reported in this paper contributes by complementing existing research to provide a deeper understanding of the status quo and challenges in engineering ML-enabled systems. In particular, we provide additional and detailed insights into contemporary practices and main problems as reported by a large sample of practitioners.